# Western Mediterranean wetlands bird species classification: evaluating small-footprint deep learning approaches on a new annotated dataset

Juan Gómez-Gómez, Ester Vidaña-Vila*, Xavier Sevillano

*GTM - Grup de recerca en Tecnologies Mèdia, La Salle-Universitat Ramon Llull, Quatre Camins, 30, 08022 Barcelona, Spain*

**Abstract**

The deployment of an expert system running over a wireless acoustic sensors network made up of bioacoustic monitoring devices that recognise bird species from their sounds would enable the automation of many tasks of ecological value, including the analysis of bird population composition or the detection of endangered species in areas of environmental interest. Endowing these devices with accurate audio classification capabilities is possible thanks to the latest advances in artificial intelligence, among which deep learning techniques excel. However, a key issue to make bioacoustic devices affordable is the use of small footprint deep neural networks that can be embedded in resource and battery constrained hardware platforms. For this reason, this work presents a critical comparative analysis between two heavy and large footprint deep neural networks (VGG16 and ResNet50) and a lightweight alternative, MobileNetV2. Our experimental results reveal that MobileNetV2 achieves an average F1-score less than a 5% lower than ResNet50 (0.789 vs. 0.834), performing better than VGG16 with a footprint size nearly 40 times smaller. Moreover, to compare the models, we have created and made public the Western Mediterranean Wetland Birds dataset, consisting of 201.6 minutes and 5,795 audio excerpts of 20 endemic bird species of the Aiguamolls de l'Empordà Natural Park.

## 1. Introduction

Bioacoustic avian life monitoring is a valuable means for obtaining relevant information regarding birdlife in a specific environment (Shonfield and Bayne 2017).

In this sense, the deployment of expert systems running over bioacoustic monitoring devices in areas of environmental interest paves the way for remote bioacoustic sensing (Wijers et al. 2021). This enables the study of birdlife in little accessible locations (e.g. reedbeds, or in strictly protected areas), or when access becomes more complicated (for instance, at night, during wintertime in highlands, or during lockdown periods).

For instance, bioacoustic avian data can be used for the quantitative and qualitative analysis of bird population composition (Rosenstock et al. 2002; Klingbail and Willig 2015), the study of different ecological moments of a species (migration or reproductive seasons), or the detection of the presence of endangered species in a particular site (Garnett et al. 2011).

However, the analysis of the recordings captured by acoustic monitoring devices is still often made by human experts (e.g. through the visual inspection of sonograms), which turns out to be a complex and time-consuming task.

Fortunately, the advances in audio signal processing and artificial intelligence have enabled the development of algorithms capable of detecting bird sounds from the recordings (Tseng et al. 2020), (Stowell et al. 2019) and classifying bird species upon their sounds (see e.g. (Knight et al. 2020)), which eases the automation of these relevant birdlife monitoring tasks.

*Corresponding author at: GTM - Grup de recerca en Tecnologies Mèdia, La Salle-Universitat Ramon Llull, Quatre Camins, 30, 08022 Barcelona, Spain

*Email addresses:* `juan.g@students.salle.url.edu` (Juan Gómez-Gómez), `ester.vidana@salle.url.edu` (Ester Vidaña-Vila ), `xavier.sevillano@salle.url.edu` (Xavier Sevillano)

In this context, the literature reflects a recent shift in the type of techniques employed for sound-based bird species classification. Indeed, early efforts in this area typically applied one-dimensional features already employed for human speech recognition (Franzen and Gu 2003) (e.g. wavelets (Selin et al. 2006), Mel-frequency cepstrum coefficients (Murcia and Paniagua 2013), hidden Markov models (Jančovič and Köküer 2016) or combinations of multiple one-dimensional features (Vidaña-Vila et al. 2020)). However, recent works increasingly use two-dimensional audio representations based on spectrograms (thus capturing the temporal evolution of audio frequency contents) (Briggs et al. 2012; de Oliveira et al. 2015; Priyadarshani et al. 2018) combined with deep neural networks (which excel in image classification tasks (Krizhevsky et al. 2017)) for audio-based bird species recognition.

This approach has the advantage of automatically learning features directly from the data during the training process, eliminating the need of using hand-engineered features. Moreover, it has been shown that image-inspired methods outperform signal-descriptive methods, which is a proof of the efficiency of spectrograms as sound descriptors (Sprengel et al. 2016; Chandu et al. 2020; Florentin et al. 2020).

However, ecologists willing to apply deep learning classification techniques for bioacoustic monitoring should take into account some issues.

First, properly training a deep neural network from scratch requires huge amounts of annotated audio recordings. In fact, their performance increases logarithmically with the volume of properly annotated training data (Sun et al. 2017). Whereas there exist several excellent and publicly available repositories of bird sounds recordings (e.g. Xeno-canto (Xeno-canto 2021) or Avibase (Lepage 2021)), these often suffer from label reliability and the presence of environmental noise (Vidaña-Vila et al. 2017), requiring further annotation efforts to provide training data for the classifiers to perform properly. Thus, the lack of annotated bird song audio data collections constitutes a bottleneck in the development of this kind of approaches.

And second, deep neural network models typically consist of millions of parameters, often reaching footprint sizes of hundreds of Megabytes, which hinders their embedding in affordable bioacoustic monitoring devices. For this reason, a standard approach is the deployment of 'dumb' sensors that just send the data to servers in the cloud, which is where the 'intelligence' resides. However, in an environmental monitoring context, access to the cloud cannot always be assumed nor guaranteed, especially in remote areas. For this reason, the *edge computing* and *fog computing* paradigms applied to deploy wireless acoustic sensor networks (i.e. performing all the computation in the node) are becoming hot research trends, which lead researchers to investigate in small-footprint algorithms that can be embedded in low-cost devices (Vidaña-Vila et al. 2020).

Taking all these considerations into account, in this work we propose to explore the performance of small-footprint deep neural networks for audio-based bird species classification, which would be a first step towards the development of affordable bioacoustic avian life monitoring devices with classification capabilities. To that end, we present a critical comparison of the performance of three deep neural network models originally pre-trained for image classification tasks, which have been subsequently fine-tuned to classify bird species from the spectrogram images of audio clips.

To tackle this fine-tuning process, we have created a brand new annotated bird song audio dataset of species belonging to the authors closest environment, the Western Mediterranean coast. To that end, we took as a reference scenario the Aiguamolls de l'Emporda` Natural Park, located in Catalonia, in north eastern Spain (42° 13' 28.09" N, 3° 05' 34.92" E) (Rosell et al. 2013). This Mediterranean coastal wetland covers 4729 ha, and its privileged location between the Muga and Fluvia` rivers makes it a natural shelter and rest area for over 320 nesting and migratory bird species, 82 of which have been reported to be regular nesters (Fatorić and Morén-Alegret 2013). Advised by biologists of the park, 20 of its endemic species were selected. Using recordings of these species available in the Xeno-Canto portal, we collected and annotated a total 5,795 clips totalizing over 200 hours of bird songs.

Both the new annotated Western Mediterranean Wetland Birds audio dataset and the fine-tuned deep neural network models are made available to the community.

## 2. Materials and methods

### 2.1. Bird song dataset

Bird data acquisition and annotation is typically an exhaustive and time-consuming task, which requires expert birders and birdwatchers to manually annotate audio files. In this regard, several platforms such as Xeno-Canto (Xeno-canto 2021) present an excellent alternative for content sharing, allowing the scientific community to have access to thousands of bird audio files. However, as the Xeno-Canto portal has not been conceived as a database for training machine learning or deep learning algorithms, when using their data for that purpose, the audio-files require a pre-processing stage to remove background noise and sounds that may belong to sources different than the bird of interest.

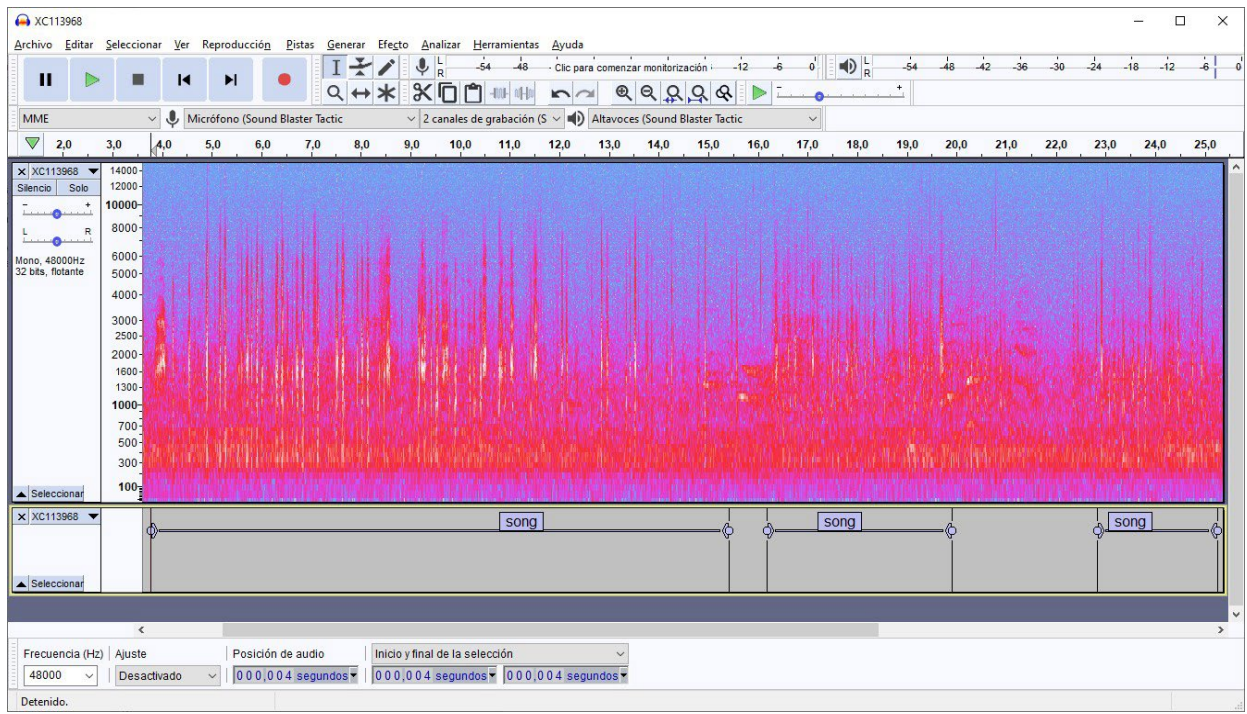

Figure 1: Screenshot of Audacity, the framework used for manually labelling the audio files.

*2.1.1. Data gathering*

As the aim of this project was to build a bird species recognition system using deep learning algorithms, the first step of the project was to collect recordings from Xeno-Canto corresponding to the following 20 endemic bird species of interest of the Aiguamolls de l'Emporda` Natural Park: *Acrocephalus arundinaceus, Acrocephalus melanopogon, Acrocephalus scirpaceus, Alcedo atthis, Anas strepera, Anas platyrhynchos, Ardea purpurea, Botaurus stellaris, Charadrius alexandrinus, Ciconia ciconia, Circus aeruginosus, Coracias garrulus, Dryobates (Dendrocopos) minor, Fulica atra, Gallinula chloropus, Himantopus himantopus, Ixobrychus minutus, Motacilla flava, Porphyrio porphyrio* and *Tachybaptus ruficollis*.

The quality of the audio files is especially relevant when deciding what audios to include in the dataset. In Xeno-Canto, audios are labelled by quality from A to E, where A means that the audio quality is excellent and E means that the audio quality is poor.

Hence, authors gathered only files that were labelled with categories A and B from the selected species. This ensured the exclusion of audios with high background noise, audios in which the bird is only heard in the background or audios with presence of other interfering sounds.

*2.1.2. Data pre-processing*

The data pre-processing task was manually carried out using the Audacity software (https://www.audacityteam.org/), a free open source program that allows to listen, record and easily label audio files.

Figure 1 depicts a screenshot of Audacity. As it can be seen, the process of labelling consists on opening a complete audio file in one track (pink and yellow spectrogram) and then creating an empty labels track. The person dedicated to label must then carefully listen to the audio file and mark in the labels track which region of the file contains relevant information. In this case, the relevant information was given the label 'song'. After labelling the full audio file, a labels file is exported in .txt format with the same name as the audio file (the Xeno-Canto identifier number). Figure 2 shows an example of a labels file for the audio file with the Xeno-Canto identifier XC127375. Notice that the file is organised in three columns: the first one indicates the starting time (in seconds) of the bird song, the second column indicates the ending time (in seconds) of the bird song, and the third column is the label of the event. In this case, the label is 'song'. The seconds are always relative to the start of the original Xeno-Canto file.

Figure 2: Example of a labels file after the labelling stage.

*2.1.3. Final dataset*

After the manual labelling process, the authors obtained a dataset of **201.6 minutes** (12,096 seconds) and **5,795 audio excerpts**. Table 1 shows in detail the amount of acoustic information obtained per each species. In most of the species, only call vocalizations or song vocalization have been considered. However, for some species, both calls and songs have been obtained. This decision was taken depending on the amount of available samples on the Xeno-Canto portal. Moreover, for the *Dendrocopos minor* species, the *drumming* effect has been selected as the identifier of the presence of the bird, despite of being a sound that is characteristic for all woodpeckers in general, and not only the *Dendrocopos minor*. The reasoning of this decision is that in the area where birds want to be surveyed is typically inhabited by the *Dendrocopos minor* species, and not other woodpeckers. A similar case occurs for the species *Ciconia ciconia*, that produces a characteristic sound produced by a repetitive clap with their bills that has been labelled as *bill clapping*. It is acoustically similar to the drumming of woodpeckers, but the spectral distribution is different as woodpeckers drum on trees using their bill and *Ciconia ciclonia* specimens use exclusively their bills to clap.

*2.2. Audio feature extraction*

To characterise the audio events so they can be automatically classified, the spectrograms of the audio files have been generated. A spectrogram is a representation of an acoustic signal. Concretely, it plots the variation of frequency of a signal in function of the variation of intensity on time: the variation on intensity is represented as a change on the brightness or color of the signal (like in a heat map), the variation on frequency is plotted vertically and the time evolution is displayed horizontally. This way, the spectrogram of an audio fragment can be regarded as a 'picture' of an acoustic event. In this work, the spectrograms have been calculated using 1-second windows, meaning that the manually cleaned audio files of the dataset were split in fragments of one second each. For those audio files shorter than one second, and as usually done in these type of problems (Singh et al. 2019), the vocalization was repeated to fill the window size. The reasoning behind choosing a window size of one second is that the window should contain—at least—one complete bird vocalization so a pattern can be obtained from the spectrogram. Also, the window should contain the fewest possible amount of noise. However, the duration of the audio files on the dataset for each species are very different (see Figure 3), which makes it hard to choose a window size that satisfies both requirements (one complete bird sound and without background noise). For this reason, the selected window offers a fair trade-off between both restrictions. Whereas in some specie, such as *Acrocephalus arundinaceus, Acrocephalus melanopogon* or *Acrocephalus scirpaceus* a single audio file will be divided in several windows, in some others such as *Ixobrychus minutus* it is very likely that the audio has to be repeated in order to fill the window. Also, the mel-scale has been used to emphasise the frequencies of interest, obtaining at the end what is typically referred to as the mel-spectrogam of the audio file.

| Bird species | Sound type | Total time (sec.) | Number of cuts |
|---|---|---|---|
| *Acrocephalus arundinaceus* | songs | 1982 | 453 |
| *Acrocephalus melanopogon* | songs | 2037 | 221 |
| *Acrocephalus scirpaceus* | songs | 2360 | 121 |
| *Alcedo atthis* | songs and calls | 351 | 418 |
| *Anas strepera* | songs | 292 | 96 |
| *Anas platyrhynchos* | songs | 229 | 70 |
| *Ardea purpurea* | calls | 128 | 207 |
| *Botaurus stellaris* | songs | 414 | 436 |
| *Charadrius alexandrinus* | songs and calls | 109 | 375 |
| *Ciconia ciconia* | bill clapping | 479 | 121 |
| *Circus aeruginosus* | calls | 185 | 307 |
| *Coracias garrulus* | calls | 178 | 267 |
| *Dendrocopos minor* | drumming | 563 | 494 |
| *Fulica atra* | calls | 123 | 372 |
| *Gallinula chloropus* | calls | 107 | 262 |
| *Himantopus himantopus* | calls | 1212 | 277 |
| *Ixobrychus minutus* | songs and calls | 148 | 559 |
| *Motacilla flava* | songs | 292 | 400 |
| *Porphyrio porphyrio* | songs and calls | 363 | 186 |
| *Tachybaptus ruficollis* | songs | 543 | 153 |
| **Total** | - | 12,096 | 5,795 |

Table 1: Length of the Western Mediterranean Wetland Birds dataset in terms of time (seconds) per species and number of audio files per species after the manual labelling process.

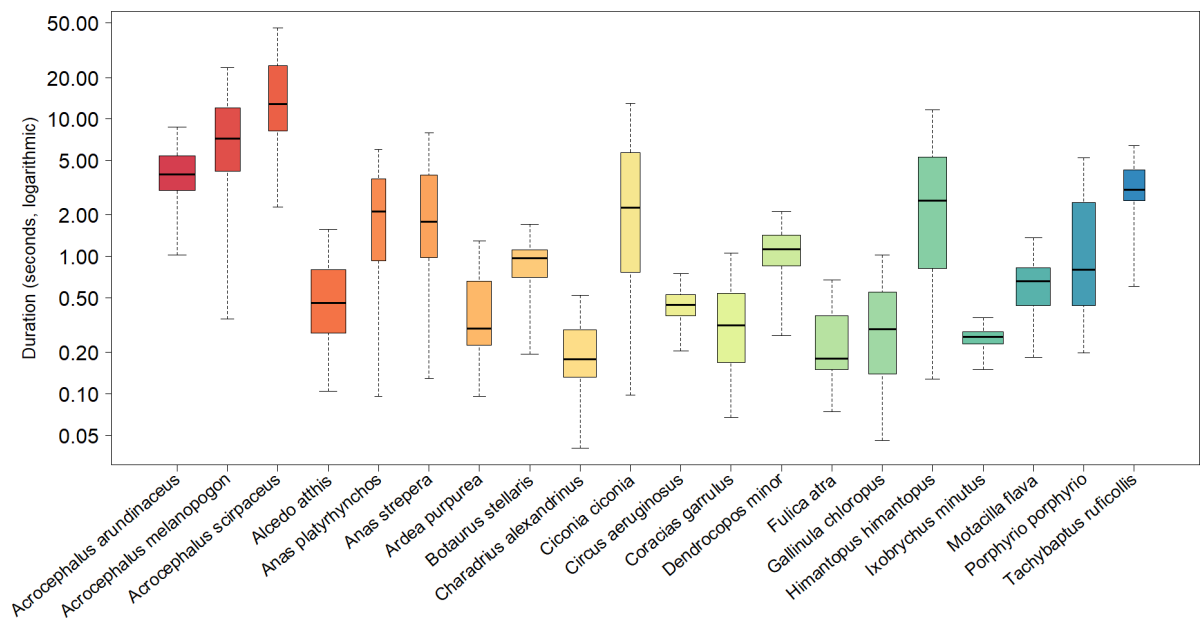

Figure 3: Boxplot representing the duration in seconds of each audio fragment of the Western Mediterranean Wetland Birds dataset, divided per species. The width of each bar represents the amount of audio fragments from that species in the dataset.

Figure 4 shows an example spectrogram for each bird species of the Western Mediterranean Wetland Birds dataset. As it can be seen, spectrograms are typically different for each bird species and follow different patterns, making these representations suitable to train machine learning or deep learning models. All the spectrograms of the Figure represent a 1-second window, and the displayed frecuencies range from 0 Hz to 11,025 Hz (sampling frecuency of 22,050 Hz).

*2.3. Classification*

Once the audio files have been converted into spectrograms, we can approach the recognition of bird species from sound as an image classification problem using a type of deep neural network called Convolutional Neural Networks (CNNs).

Whereas CNNs are typically used for image classification tasks, the literature (Knight et al. 2020; Florentin et al. 2020; Sprengel et al. 2016) has demonstrated that high accuracy bird species audio classification can be achieved for different species and types of vocalisation by using spectrograms.

As mentioned earlier, one of our goals is to evaluate the performance of small-footprint networks, as a first step towards the development of affordable and 'intelligent' bioacoustic monitoring devices.

For this reason, in this work we make a critical comparison between three well-known CNNs, to evaluate the effect of network architecture, depth and size on the overall classification accuracy. The three models under comparison are: *i)* VGG16 (Simonyan and Zisserman 2014), a relatively shallow deep network *ii)* ResNet50 (He et al. 2016), a very deep and complex network and *iii)* MobileNetV2 (Howard et al. 2017), a small-footprint network designed for being embedded in mobile devices, which can be of interest for developing affordable bioacoustic monitoring devices.

In the following sections, we summarise the architectures of the evaluated CNNs, describe their training and test processes, as well as the metrics employed for evaluating their performance.

*2.3.1. Network architecture*

Any traditional CNN architecture can be divided into two distinct parts: convolutional layers and fully-connected layers, with the latter placed on top of the former. The last fully-connected layer uses the softmax activation function and a number of neurons equal to the number of classes we intend to classify, to produce a vector of probabilities corresponding to the class predictions.

The VGG16 model only uses 3x3 convolutional layers stacked on top of each other, reaching a maximum of 16. To reduce the volume size, max pooling is used. At the top, there are two fully-connected layers followed by a softmax classifier. The full VGG16 network architecture is shown in Figure 5.

The ResNet50 model relies on a micro-architecture called residual module which allows to train very deep CNN networks without stagnating while training. It has 48 convolutional layers, followed by a mean clustering layer and a dense layer that serves as the output of the network to obtain probabilities. Its architecture is presented in Figure 6a.

Finally, MobileNetV2 architecture is based on a streamlined architecture that uses depth-wise separable convolutions to build lightweight deep neural networks. An interesting feature of this CNN is that by tuning of two of its hyper-parameters, it is possible to balance the size model having speed and accuracy as trade-offs. The MobileNetV2 architecture is depicted in Figure 6b.

In this work, the implementations of the three CNNs employed are obtained directly from the Applications API of the Keras deep learning library (Chollet et al. 2015). A summary of the most important characteristics of the three models is presented in Table 2, including the model footprint size in Megabytes, the number of tunable parameters and the topological depth. Notice how MobileNetV2 has a much smaller footprint despite being a much deeper network.

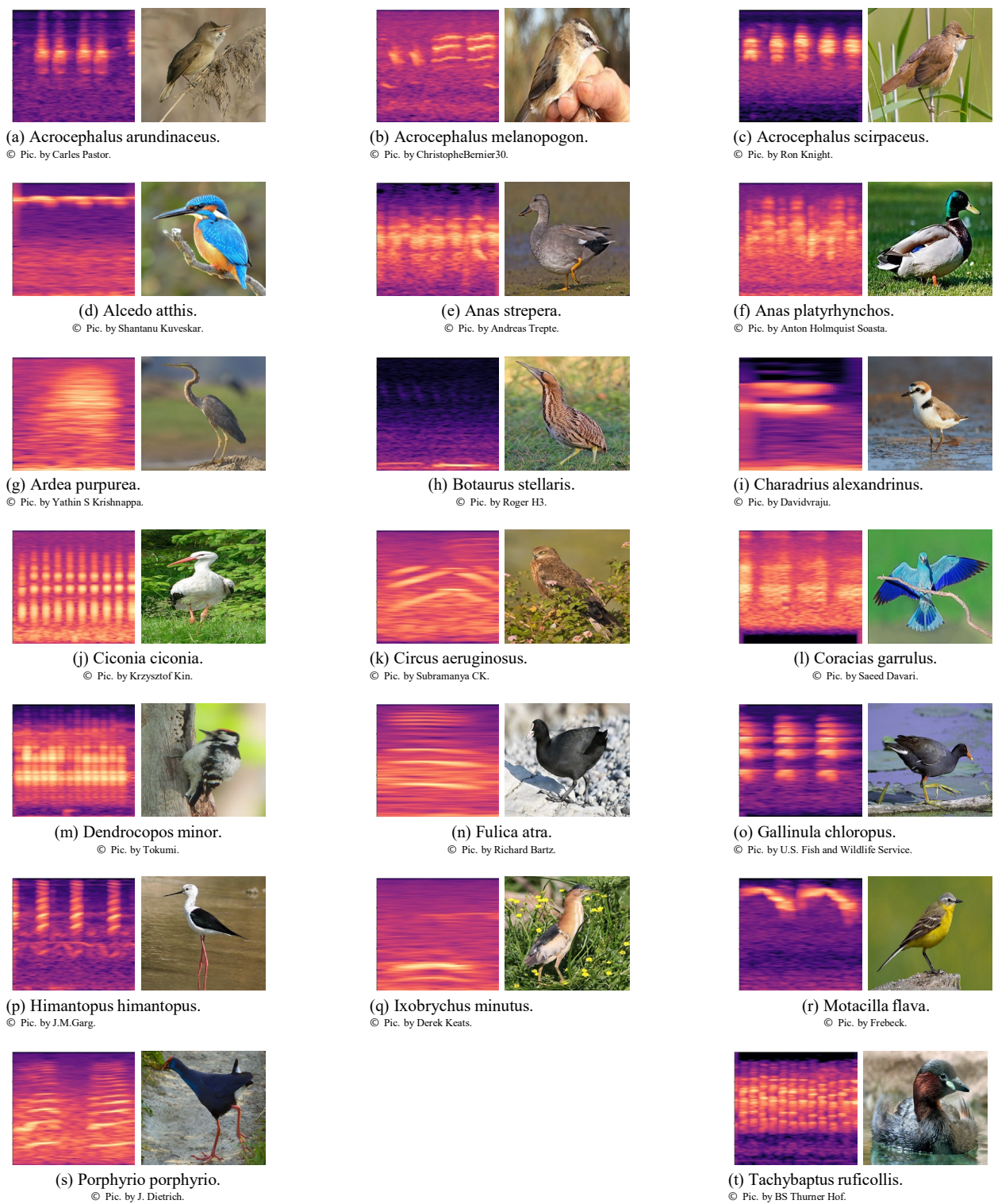

Figure 4: Pictures of the birds of the Western Mediterranean Wetland Birds dataset and an example spectrogram of their vocalizations for each species.

| Model | Size | Parameters | Depth |
|---|---|---|---|
| VGG16 | 528 MB | 138,357,544 | 23 |
| ResNet50 | 98 MB | 25,636,712 | 50 |
| MobileNetV2 | 14 MB | 3,538,984 | 88 |

Table 2: Summary of the architecture features of the VGG16, ResNet50 and MobileNetV2 CNNs.

*2.3.2. Network training and testing*

In this work, we have followed a fine-tuning strategy to train the CNNs, an approach that takes advantage of using an already validated and pre-trained network architecture. Moreover, it also allows to obtain high classification accuracies without the need of compiling a dataset of thousands of spectrogram images per class, while avoiding long and complex training processes.

Fine-tuning consists on replacing the top fully-connected layers of already pre-trained CNN models with new ones initialised with randomised weights. This allows to train the neural networks to classify the classes of our choice rather than those predefined by prior learning.

In our case, the VGG16, ResNet50 and MobileNetV2 models already pre-trained on the ImageNet data set (Deng et al. 2009) (which are available in the Applications API of Keras (Chollet et al. 2015)) have been employed.

As the convolutional layers of the pre-trained models have already learned and tuned filters, fine-tuning is performed by freezing all layers in the body of the network and then training only the new fully-connected head as a warm-up phase using part of the spectrogram dataset described in Section 2.1. After this first phase, the original layers are unfrozen and another training phase with a very small learning rate is carried out to increase the overall accuracy.

As for the tuning of the network hyperparameters, the values of some of them are duplicated in both training phases. However, others are independently fixed for each phase, and they are presented in Table 3. These values are the result of multiple experiments and tuning that have led to the maximum accuracy without long training times or overfitting. Notice that the number of training epochs for the warm-up phase is between 20 and 30, to give enough time to train the parameters of the new top layers of the network. It is also worth mentioning that between the two fully-connected top layers of the architectures of all three models, a dropout (Srivastava et al. 2014) with a rate of 0.5 has been employed to provide regularisation.

As for the input data that is fed into the CNNs, it must have a size of (X, Y, 3), meaning it needs three input channels. As spectrograms are only 2D, and in our case of size (224, 224), we need to triplicate the layers into the third dimension, thus obtaining an input size of (224, 224, 3) for each spectrogram image.

Finally, we have split our dataset into three slices of 70%/20%/10% for training, validation and test respectively. All spectrograms coming from the same audio file will only belong to a specific subset to avoid giving our neural network model an advantage when classifying spectrograms outside the training stage, which would result in a falsely improved accuracy score.

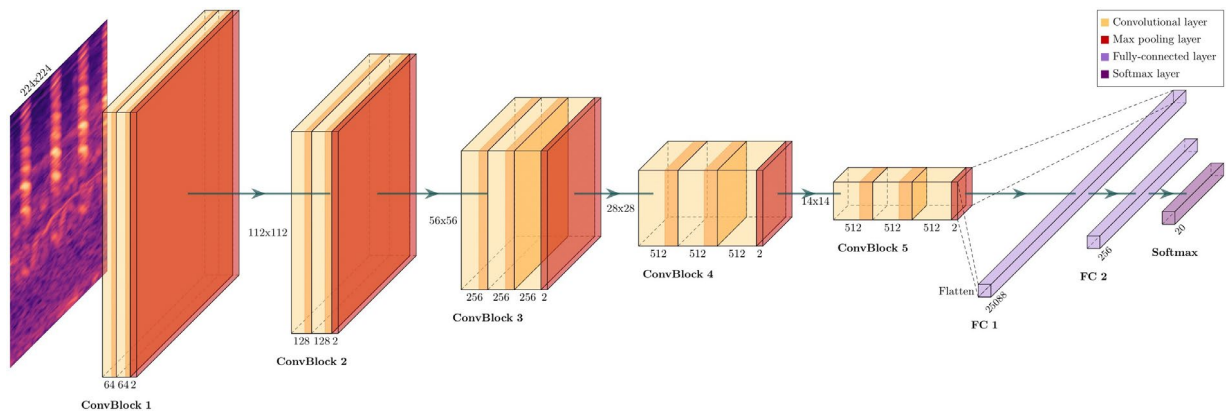

Figure 5: VGG16 network architecture with custom top layers.

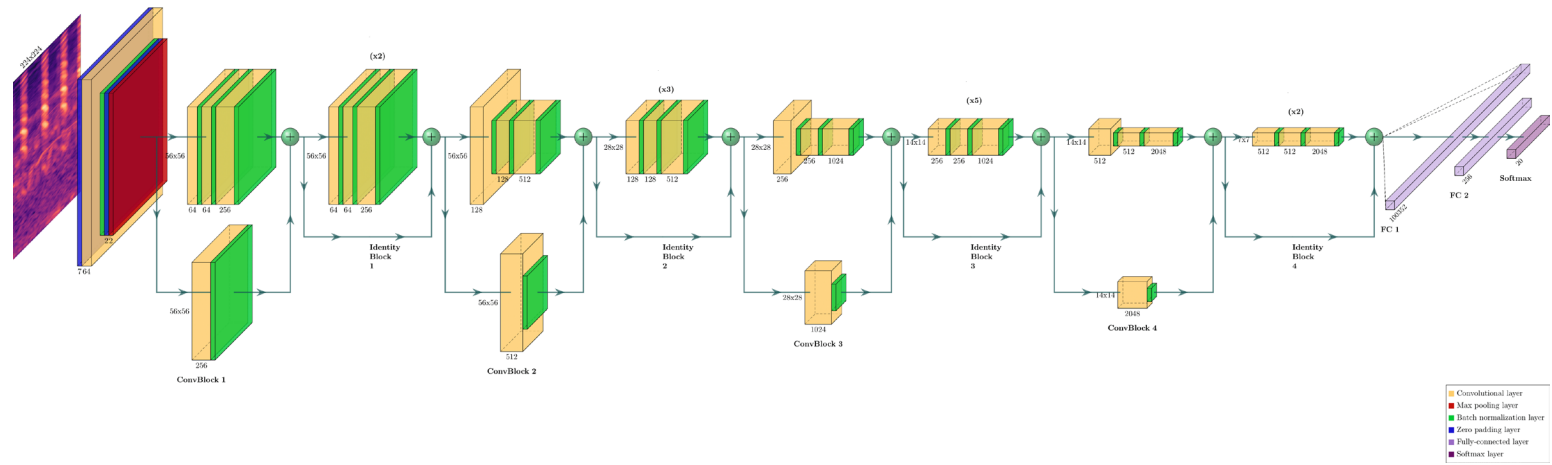

(a) ResNet50 architecture

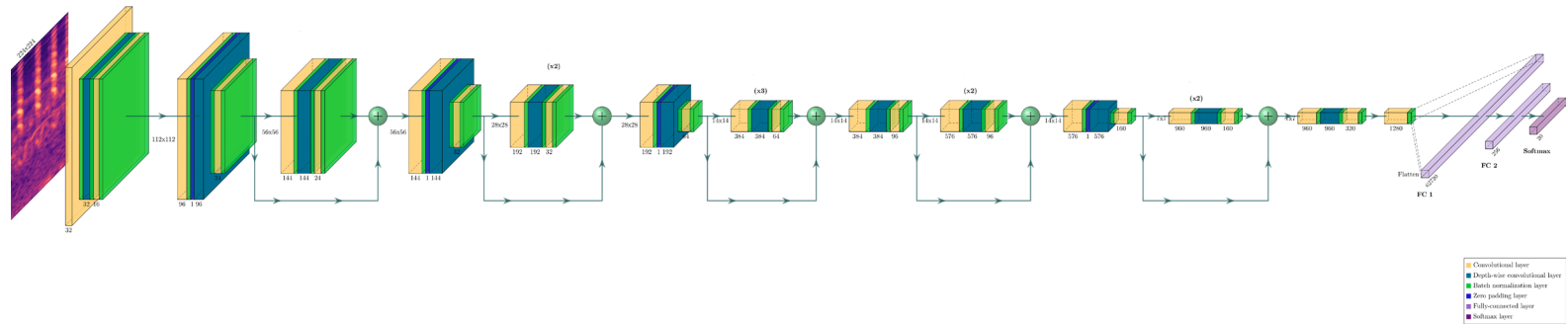

(b) MobileNetV2 architecture

Figure 6: ResNet50 (top) and MobileNetV2 (bottom) network architectures with custom top layers.

| Hyperparameter | Value |
| --- | --- |
| # Epochs 1 (warm-up phase) | 25 |
| # Epochs 2 | 50 |
| Batch size | 16 |
| Optimizer 1 (warm-up phase) | RMSprop |
| Learning rate 1 (warm-up phase) | 0.001 |
| Optimizer 2 | SGD |
| Learning rate 2 | 0.001 |

Table 3: Training hyperparameters for all three models.

*2.3.3. Evaluation metrics*

To provide a comprehensive view of the performance of the deep neural network models, they have been evaluated with three metrics: precision, recall and F1-measure.

These evaluation metrics are obtained from the classifier *confusion matrix*, which offers a clear view of which classes are more and less successfully classified. This is because on a confusion matrix, all classes are represented on the columns. On the rows, the same classes are presented but in a predictive form.

So, for class *i*, the element placed on the *i*th row and *i*th column (that is, on the diagonal) corresponds to the correct predictions score (*true positive* rate or TP). The other cells of the diagonal represent all the correct negative predictions (*true negative* rate or TN). The off-diagonal confusion matrix cells represent the two possible types of prediction errors: *false positives* (FP) are on the *i*th row, and represent incorrect assignments of instances of other classes to class *i*, while *false negatives* (FN) appear on the *i*th column, representing cases when the model wrongly assigned instances of class *i* to other classes.

With these four values, three key performance metrics are calculated for each class, with scores ranging from 0 to 1, being 1 the perfect score.

- *Recall* (or sensitivity) represents the fraction of the instances of class *i* that were correctly classified:

$$R = \frac{TP}{TP + FN} \qquad (1)$$

Put in bioacoustic avian monitoring terms, recall would represent the ability of the system not to miss an acoustic event of a specific species, as it represents the probability that, if a bird should be classified as class *i*, this decision is taken.

- *Precision* (or specificity) measures the fraction of real class *i* instances among those assigned to class *i*:

$$P = \frac{TP}{TP + FP} \qquad (2)$$

Thus, precision represents the probability that if a bird is classified as class *i*, this decision is correct.

- *F1-score* is the harmonic mean of precision and recall:

$$F_1 = \frac{2 \cdot P \cdot R}{P + R} \qquad (3)$$

This score unifies these two complementary performance measures into a single metric that gives the same importance to both measures.

Finally, to obtain global scores for each classifier, the macroaveraging strategy is used, which consists in computing the aforementioned metrics for each class, and then averaging over all the classes.

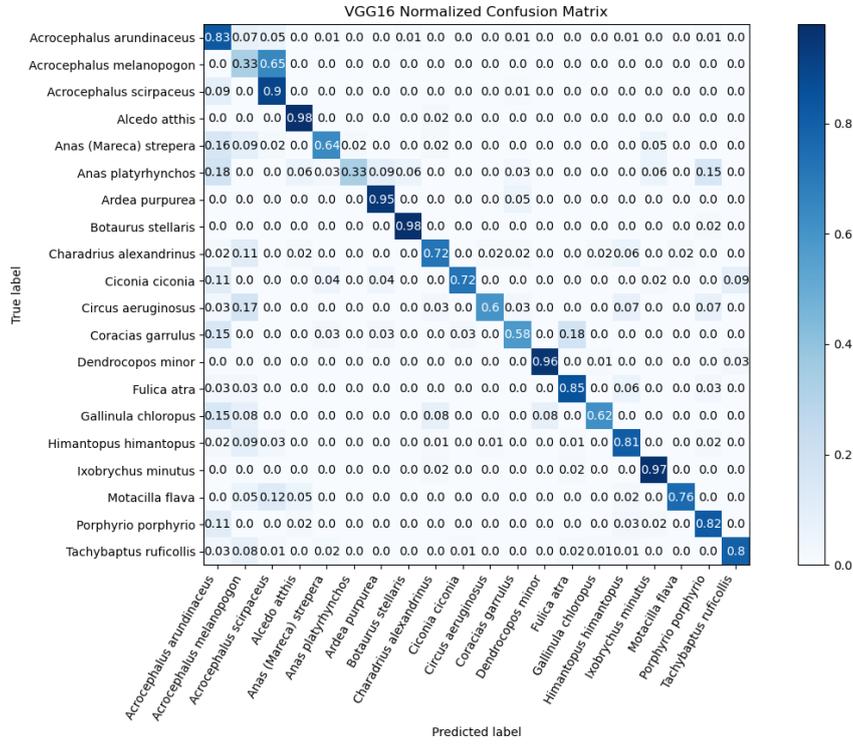

(a) VGG16 confusion matrix

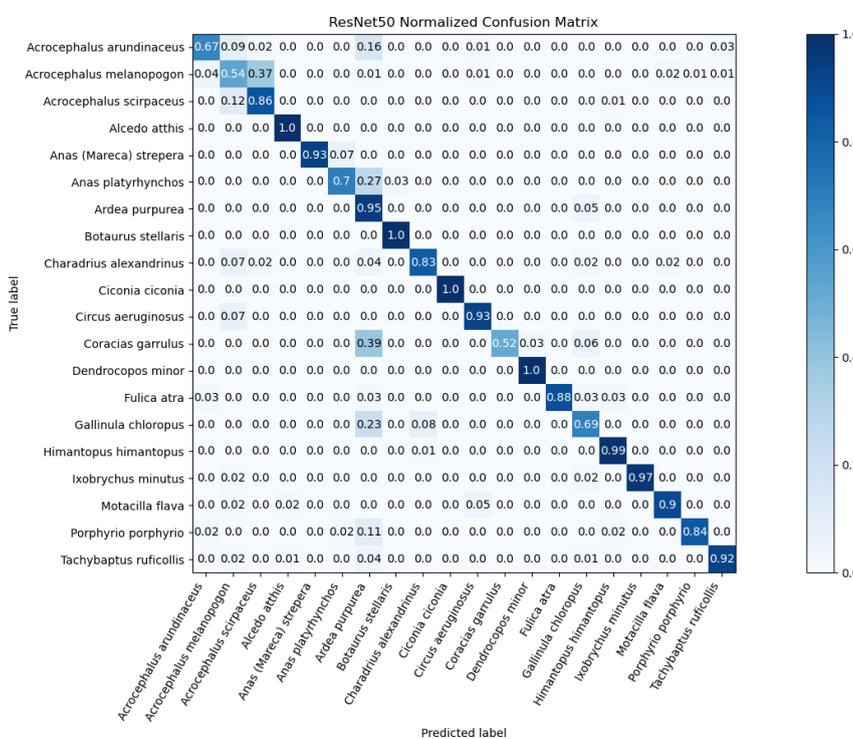

(b) ResNet50 confusion matrix

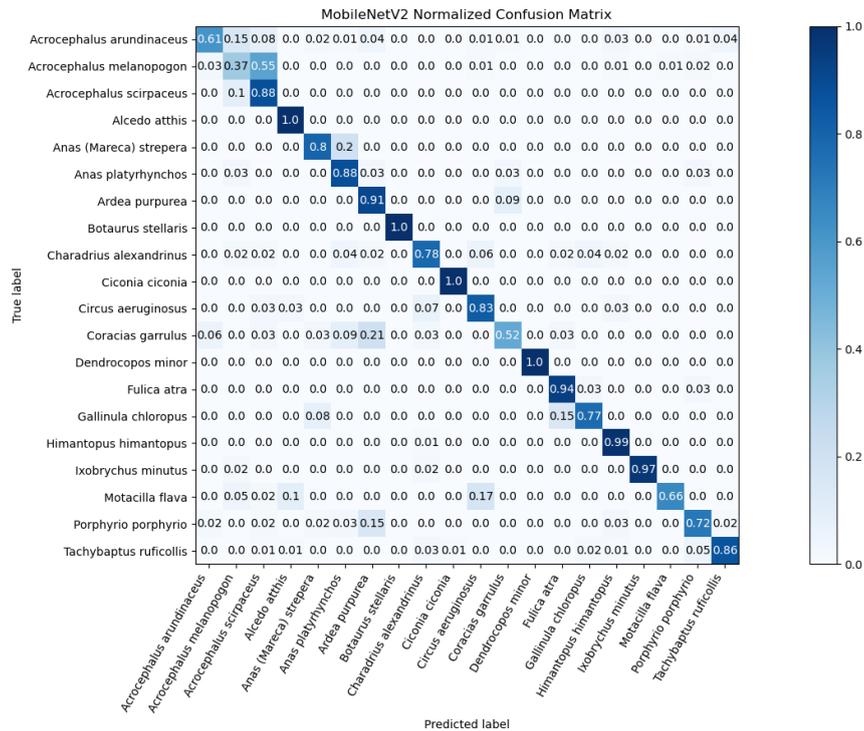

(c) MobileNetV2 confusion matrix

Figure 7: VGG16 (a), ResNet50 (b) and MobileNetV2 (c) normalised confusion matrices.

## 3. Results and discussion

### 3.1. The Western Mediterranean Wetland Birds dataset

The first outcome of this work is the annotated Western Mediterranean Wetland Birds dataset, which is available to the scientific community in two forms.

First, we share the total of 5,795 annotated audio clips generated from a source of 1,098 recordings, adding up to a total of 201.6 minutes (12,096 seconds) of vocalizations of different lengths, alongside with their corresponding annotations.

And second, we also share the Mel spectrogram version of the dataset, where each image corresponds to 1-second window of the original audio, resulting in a total of 17,536 spectrogram images stored in matrix form.

These two versions of this brand new annotated bird song dataset are available for download at https://doi.org/10.5281/zenodo.5093173.

### 3.2. Classifier performance analysis

All three VGG16, ResNet50 and MobileNetV2 models have been trained and tested using the methodology described in Section 2.3.2. As with the dataset, the models are also available at https://github.com/jogomez97/BirdMLClassification.

Figure 7 depicts the confusion matrices for the three models, which is used as a reference for at-a-glance analysis of their performance. Next, the analysis of each individual model performance is presented, including an analysis of the bird species confusion. To conclude, we present a global comparison between models.

#### 3.2.1. VGG16 model

The VGG16 network architecture reaches a macroaveraged recall of 0.757, a precision of 0.812 and a F1-score of 0.768.

In general, the model predicts most of the classes with high accuracy, as it can be observed in the confusion matrix of Figure 7a.

Out of the 20 classes, 13 have an F1-score greater than the average value of 0.768, while 18 have a performance higher than 0.5.

The remaining two classes have lower F1-scores: 0.43 for *Acrocephalus melanopogon* and 0.49 for *Anas platyrhynchos*.

As for the former, the inspection of the confusion matrix in Figure 7a shows that VGG16 mistakes *Acrocephalus melanopogon* for *Acrocephalus scirpaceus*, classifying two thirds of the samples of the first species as the second. In reality, both these species have similar songs, which can be considered the reason of this confusion.

In the case of *Anas platyrhynchos*, we observe more diversity in its errors. Considering that VGG16 managed to achieve high classification accuracy on other bird species with similar amounts of data, we conjecture that in this case the model has not been able to generalise from the spectrograms of the species.

*3.2.2. ResNet50 model*

The classification performance of ResNet50 is far superior than VGG16, achieving macroaveraged recall and precision scores of 0.856 and 0.855 respectively, and a F1-score of 0.834.

With the ResNet50 model, 13 out of 20 classes have a F1-score higher than the average 0.834, and 19 classes score above the 0.5 mark, which represents a significant improvement with respect the VGG16 model.

Focusing on the two bird species that were less well classified by VGG16, we observe that *Anas platyrhynchos* significantly increases its F1-score from 0.49 to 0.77. This reinforces the notion that the VGG16 model was unable to generalise this class, while the more complex ResNet50 architecture is able to do so.

As for *Acrocephalus melanopogon*, its macroaveraged F1-score increases from 0.43 to 0.61. Despite this clear improvement, we still observe there is a certain degree of confusion between species *Acrocephalus melanopogon* and *Acrocephalus scirpaceus*.

For the ResNet50 model, the class with the lowest F1 score is *Ardea purpurea*, which achieves a 0.30 compared to the 0.82 of the VGG16 model. This is because for this class, a recall of 0.95 is reached (i.e. the model rarely misses events of this species), but precision is only 0.18 (i.e. it tends to confuse other species with it), resulting in a low F1-score. We consider this may be due to the small number of samples of this class, which makes any classification error of the model to have a large impact on the value of precision.

*3.2.3. MobileNetV2 model*

Despite being the one with the smallest footprint, the MobileNetV2 model seems to achieve a trade-off performance between the two previous networks, as it achieves a macroaveraged recall of 0.824, a precision of 0.785 and a F1-score of 0.789.

With the MobileNetV2 model, 10 classes have a F1-score higher than the average 0.789, and 19 achieve a F1-score higher than 0.5.

What is truly remarkable, and specially for a network as small as MobileNetV2, is that overall, there is only one class that has a poor F1-score. That is the case of *Acrocephalus melanopogon* species with a 0.45 score (0.43 and 0.61 for VGG16 and ResNet50 models, respectively). This shows that this particular model has the most balanced scores across all classes.

| Model | Recall | Precision | F1-score |
|---|---|---|---|
| VGG16 | 0.757 | 0.812 | 0.768 |
| ResNet50 | 0.856 | 0.855 | 0.834 |
| MobileNetV2 | 0.824 | 0.785 | 0.789 |

Table 4: Comparison of the performance of the VGG16, ResNet50 and MobileNetV2 models on the test data in terms of macroaveraged recall, precision and F1-score.

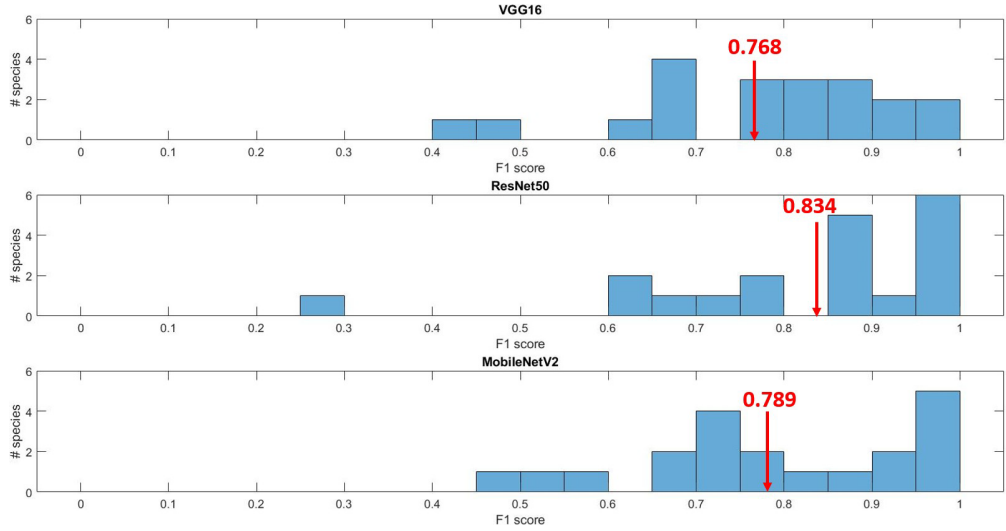

Figure 8: F1-score histograms for the VGG16, ResNet50 and MobileNetV2 models. The macroaveraged F1-score of each model is indicated in red.

*3.2.4. Summary of classifier performance*

To provide the reader with a comparison between the three evaluated models, Table 4 presents their macroaveraged recall, precision and F1-score values.

The ResNet50 model is the best performer, achieving the highest values in all three metrics, followed by MobileNetV2 and VGG16, in that order.

Focusing on the smallest footprint network, MobileNetV2, we can see that it achieves a high recall, close to that of ResNet50. This means that this network is unlikely to miss samples of a given bird species.

In contrast, its precision is a bit lower than that of its competitors. The reason for this is the confusion in the classification of the birds of the *Acrocephalus* genre (*Acrocephalus arundinaceus*, *Acrocephalus melanopogon* and *Acrocephalus scirpaceus*). Notice that the three models struggle with these three species, and MobileNetV2 happens to be the one that classifies them with the lowest precision. Instead, it is capable of classifying the remaining species with high precision, being comparable to the other two networks.

To provide the reader with a higher level comparative analysis between the classification performance of VGG16, ResNet50 and MobileNetV2, Figure 8 depicts the F1-score histograms of the 20 classes for the three models, showing their distribution.

It can be observed that MobileNetV2 has the most compact histogram. This indicates that its performance is the most stable and balanced of the three networks in the test bed, which is an interesting feature of a classifier in terms of robustness and reliability.

## 4. Conclusion

In this work we have presented a critical comparative analysis of the performance of three deep learning models on the task of bird species classification on the brand new Western Mediterranean Wetland Birds dataset.

The application of deep neural networks for bird song classification using spectrograms has been recently explored in several works (Joly et al. 2018). For instance, high classification accuracies have been achieved using well-established deep networks such as AlexNet (Chandu et al. 2020; Knight et al. 2020; Florentin et al. 2020), Inception (V3 and/or V4) (Lasseck 2018; Florentin et al. 2020; Sevilla and Glotin 2017), VGG19, different ResNet variants or DenseNet (Lasseck 2018; Florentin et al. 2020).

However, those works paid little attention to the network footprint, which ranges between 57 MB (for DenseNet) and 549 MB (for VGG19).

In our work, we have focused our analysis on whether a small footprint network like MobileNetV2 can perform comparably to heavier models, such as VGG16 or ResNet50.

In our experiments, MobileNetV2 has demonstrated to achieve a largely stable and accurate performance, close enough to the best-performing ResNet50 model despite having eight time less parameters and a model size 7 times smaller. Also, MobileNetV2 outperforms a much larger and heavier deep learning model as VGG16, despite having 97.44% less parameters.

These results demonstrate that it is possible to design affordable bioacoustic monitoring devices (i.e. resource and battery constrained) that perform reliably thanks to embedded audio classification capabilities powered by small footprint deep learning techniques. We believe this conclusion will be useful for ecologists willing to develop bird species classification systems powered by deep learning, but who are not knowledgeable in the field.

We consider this is a very appealing result that paves the way for the automation of bioacoustic avian life monitoring in scenarios where wireless access to cloud computing facilities is difficult or not possible.

The low-level analysis of the classification performance of the networks in our test bed reveals that confusions between songs of birds of the same genre are common. This issue has been recently studied in the case of calls of *Catharus bicknelli* and *Catharus minimus* (Marchal et al. 2021), which indicates this is still a challenge for automatic bird song classification systems. For this reason, in scenarios where this situation is found, it would be advisable to collect large numbers of samples of these species to properly train the network and reduce misclassifications.

Another relevant contribution of this work is the construction of the annotated Western Mediterranean Wetland Birds dataset containing sounds of 20 endemic bird species of the Aiguamolls de l'Emporda` Natural Park.

In this sense, while it is possible to find annotated birdsong datasets (Arriaga et al. 2015; Morfi et al. 2019), it is still common to find other works that have created and annotated their own datasets, but they have not made them available to the public (Knight et al. 2020).

As the limited availability of annotated data is one of the main hurdles in the field (Vidaña-Vila et al. 2017), we have decided to make the Western Mediterranean Wetland Birds dataset public and encourage the community to use it to foster research on this topic. By publishing not only the spectrograms but also the original audio files, we encourage the use and combination of different spectrogram configurations, as they have been recently proven to better classification performance (Knight et al. 2020).

As for the future work, we consider that an interesting continuation of this work would be the creation of ensembles of networks as a means to improve classification and reduce confusion between bird species of the same genre.

**Acknowledgements**

Authors would like to acknowledge Albert Brugas Riera and Sergio Romero de Tejada Martínez from the Aiguamolls de l'Emporda` Natural Park for their valuable help when defining the species of interest to be classified. Also, authors would like to acknowledge all the Xeno-Canto community and their contributors for making the creation of the dataset possible. Specially, authors would like to thank the following contributors for giving us special permission to use their recordings in this work despite having uploaded them on the Xeno-Cano portal under the terms BY-NC-ND: Anhäuser, Arnold Meijer, Bodo Sonnenburg, Chie-Jen Jerome Ko, Ding Li Yong, Eveny Luis, Fernand Deroussen (Sonothèque du MNHN), Hans Matheve, Herman van der Meer, Itziar Gutiérrez, Jacques Prevost, Jarek Matusiak, Jérémy Simar, Joost van Bruggen, Krzysztof Deoniziak, Lars Lachmann, Mandar Bhagat, Marc Anderson, Marco Dragonetti (www.birdsongs.it), Matthias, Feuersenger, Maudoc, Niels Krabbe, Patrick Franke, Peter Boesman, Piotr Szczypinski, Ruud van Beusekom.

**Data availability**

The dataset can be downloaded from the following repository https://doi.org/10.5281/zenodo.5093173.

The deep learning models can be downloaded from the following repository https://github.com/jogomez97/BirdMLClassification.